
\magnification\magstep1

\openup 1\jot

\def\hbar{\mathchar '26\mkern -9muh}

\catcode`@=11
\def\eqaltxt#1{\displ@y \tabskip 0pt
  \halign to\displaywidth {%
    \rlap{$##$}\tabskip\centering
    &\hfil$\@lign\displaystyle{##}$\tabskip\z@skip
    &$\@lign\displaystyle{{}##}$\hfil\tabskip\centering
    &\llap{$\@lign##$}\tabskip\z@skip\crcr
    #1\crcr}}
\def\eqallft#1{\displ@y \tabskip 0pt
  \halign to\displaywidth {%
    $\@lign\displaystyle {##}$\tabskip\z@skip
    &$\@lign\displaystyle{{}##}$\hfil\crcr
    #1\crcr}}
\catcode`@=12 

\def\half{{\textstyle {1 \over 2}}}

\def\pmb#1{\setbox0=\hbox{#1}  \kern-.025em\copy0\kern-\wd0
  \kern0.05em\copy0\kern-\wd0  \kern-.025em\raise.0433em\box0 }
\def\pmbh#1{\setbox0=\hbox{#1} \kern-.12em\copy0\kern-\wd0
	    \kern.12em\copy0\kern-\wd0\box0}
\def\sqr#1#2{{\vcenter{\vbox{\hrule height.#2pt
      \hbox{\vrule width.#2pt height#1pt \kern#1pt
	 \vrule width.#2pt}
      \hrule height.#2pt}}}}

\def\rchi{{\raise 2pt \hbox {$\chi$}}}
\def\rga{{\raise 2pt \hbox {$\gamma$}}}
\def\rg{{\raise 2 pt \hbox {$g$}}}

\def\({\left(}
\def\){\right)}
\def\<{\left\langle}
\def\>{\right\rangle}

\def\[{\left[}
\def\]{\right]}
\let\text=\hbox
\def\pt{\partial}

\def\rta{\rightarrow}

\def\om{\omega}
\def\ol{\overline}

\def\sig{\sigma}

\def\ti{\tilde}

\def\cH{{\cal H}}

\def\A{\hbox{$ A\kern -5.5pt / \kern +5.5pt$}}
\def\B{\hbox{$ B\kern -5.5pt / \kern +5.5pt$}}
\def\C{\hbox{$ C\kern -5.5pt / \kern +5.5pt$}}
\def\D{\hbox{$ D\kern -5.5pt / \kern +5.5pt$}}
\def\E{\hbox{$ E\kern -5.5pt / \kern +5.5pt$}}
\def\F{\hbox{$ F\kern -5.5pt / \kern +5.5pt$}}
\def\G{\hbox{$ G\kern -5.5pt / \kern +5.5pt$}}
\def\H{\hbox{$ H\kern -5.5pt / \kern +5.5pt$}}
\def\I{\hbox{$ I\kern -5.5pt / \kern +5.5pt$}}
\def\Z{\hbox{$ Z\kern -5.5pt / \kern +.5pt$}}

\hfuzz 6pt

\catcode`@=12 

\font\twelverm=cmr12
\font\twelvebf=cmbx12
\def\bigtype{\twelverm \twelvebf \baselineskip=16pt}

\rightline{DAMTP R95/19}

\vskip .2 true in
\centerline {\bigtype Is there a problem with quantum }
\centerline {\bigtype wormhole states in N=1
Supergravity?\footnote*{\sevenrm
 Based on a essay
which received a Honorable Mention in the  1995
Gravity Research Foundation Awards.}}
\vskip .1 true in
\centerline {{\rm P.V. Moniz}\footnote{**}{{\rm e-mail address:
prlvm10@amtp.cam.ac.uk}}}
\vskip .1 true in
\centerline {Department of Applied Mathematics and Theoretical Physics}
\centerline{
University of Cambridge}
\centerline{
 Silver Street, Cambridge,
CB3 9EW, UK }
\vskip .2 true in

{\leftskip = 1.5in   {\it It was
the best of times, it was the worst of times; it was the age of wisdom, it was
the age of
foolishness; it was the epoch of belief, it was
the epoch of incredulity; it was the season of Light, it was the season of
Darkness;
it was the spring of hope, it was the winter of despair; we had everything
before us, we had
nothing before us...}

{\rm C. Dickens, A Tale of Two Cities }             \par}

\vskip .2 true in
\centerline {\bf ABSTRACT}
\vskip .1 true in

{\sevenrm
The issue concerning the existence
of wormhole states in locally supersymmetric minisuperspace
models with matter is addressed.
Wormhole  states
are apparently  absent in models obtained from
the more  general theory of N=1 supergravity with supermatter.
A Hartle-Hawking type solution can be found, even though some terms (which are
scalar field
dependent)
cannot be determined in a satisfactory way.
A possible cause  is   investigated here. As far as the wormhole situation
is concerned, we argue here
that the type
of Lagrange multipliers and fermionic derivative ordering  used can make a
difference.
A proposal is made for  supersymmetric quantum wormholes to also
be invested with a Hilbert space structure, associated with
a maximal analytical extension of the corresponding
minisuperspace.}
\noindent

\magnification\magstep1

\openup 1\jot

\input mssymb
\def\hbar{\mathchar '26\mkern -9muh}

\catcode`@=11
\def\eqaltxt#1{\displ@y \tabskip 0pt
  \halign to\displaywidth {%
    \rlap{$##$}\tabskip\centering
    &\hfil$\@lign\displaystyle{##}$\tabskip\z@skip
    &$\@lign\displaystyle{{}##}$\hfil\tabskip\centering
    &\llap{$\@lign##$}\tabskip\z@skip\crcr
    #1\crcr}}
\def\eqallft#1{\displ@y \tabskip 0pt
  \halign to\displaywidth {%
    $\@lign\displaystyle {##}$\tabskip\z@skip
    &$\@lign\displaystyle{{}##}$\hfil\crcr
    #1\crcr}}
\catcode`@=12 

\def\half{{\textstyle {1 \over 2}}}

\def\pmb#1{\setbox0=\hbox{#1}  \kern-.025em\copy0\kern-\wd0
  \kern0.05em\copy0\kern-\wd0  \kern-.025em\raise.0433em\box0 }

\def\pmbh#1{\setbox0=\hbox{#1} \kern-.12em\copy0\kern-\wd0
	    \kern.12em\copy0\kern-\wd0\box0}
\def\sqr#1#2{{\vcenter{\vbox{\hrule height.#2pt
      \hbox{\vrule width.#2pt height#1pt \kern#1pt
	 \vrule width.#2pt}
      \hrule height.#2pt}}}}

\def\rchi{{\raise 2pt \hbox {$\chi$}}}
\def\rga{{\raise 2pt \hbox {$\gamma$}}}
\def\rrho{{\raise 2pt \hbox {$\rho$}}}

\def\({\left(}
\def\){\right)}
\def\<{\left\langle}
\def\>{\right\rangle}

\def\[{\left[}
\def\]{\right]}
\let\text=\hbox
\def\pt{\partial}

\def\rta{\rightarrow}

\def\om{\omega}
\def\ol{\overline}

\def\sig{\sigma}

\def\ti{\tilde}

\def\cH{{\cal H}}


 A quantum theory of gravity
constitutes one of the foremost aspirations  in
theoretical physics
[1].
The inclusion of    supersymmetry could allow
important achievements as well. Firstly, supersymmetry is an
 attractive concept  with
appealing possibilities in particle physics. The introduction of
local supersymmetry and subsquently of  supergravity  provide
an elegant gauge theory between bosons and fermions to which many hope
nature has reserved a rightful  place [2]. In fact,
N=1 supergravity is  a (Dirac) square root of gravity [3]:
physical states in the quantum theory must satisfy the
 supersymmetry constraints which
then imply with the quantum algebra that  the Hamiltonian constraints also
 to be satisfied  [3,4,5]. Secondly,
ultraviolet divergences  could be removed by
the presence of the extra symmetry [6].
Thirdly, it was suggested [7] that Planckian
effective masses  induced by wormholes
 could be eliminated with supersymmetry.

 Quite recently, some important results were achieved [8,9].
On the one hand,  addressing the question of why the existence
of a Hartle-Hawking [10]  solution for Bianchi class A models
in pure N=1 supergravity [11-15]  seemed to depend on the
homogeneity condition for the gravitino [13].
In fact, it does not and it is now possible to find a Hartle-Hawking
and wormhole [15]  solutions in the same spectrum [8].
This result  requires    the inclusion of all allowed gravitational degrees of
freedom
into the Lorentz invariant fermionic sectors of the wave function.
On the other hand, investigating why no physical states were found in
ref. [17-19] when a cosmological constant
is added (nevertheless, a   Hartle-Hawking
solution was obtained for a $k=1$ FRW model). Extending  the
framework presented in ref. [8] and using  Ashtekar variables,
it was shown in ref. [9] that the exponential of the Chern-Simons
functional constitute one case of solutions.
However, there are many other issues in supersymmetric
quantum gravity which remain unsolved. In particular,  why the minisuperspace
solutions
have no counterpart in the full theory because states
with zero (bosonic) or a finite number of fermions are not possible there [20].
A possible answer could be provided within the framework
presented in ref. [8]. Other interesting issues in supersymmetric  quantum
gravity/cosmology are:
a) obtaining conserved currents in minisuperspace from the wave function of the
universe,
$\Psi$ [21];
b) obtaining physical states in the full theory (are there any? how do they
look?) and
possibly checking
the conjecture made in [8]; c) why  there are no physical states in a locally
supersymmetric
FRW model with gauged supermatter [22] but we can find them in  a locally
supersymmetric
FRW model with   Yang-Mills fields [23].
But another problem has also been kept
without an adequate explanation: the apparent absence of
wormhole states either in some FRW [24-26] or
Bianchi IX models [27] when supermatter is included. In addition,
a Hartle-Hawking type solution can be found, even though some terms (which are
scalar field
dependent)
cannot be determined in a satisfactory way.

Classically, wormholes  join
different asymptotic regions of a Riemannian geometry.
Such solutions can  only be found when
certain types of matter fields are present
[16].
However, it  seems more natural to study quantum
wormhole states, i.e., solutions of the Wheeler-DeWitt
equation [16,28-31]. It is thought that wormholes may produce
 shifts  in  effective masses and interaction parameters [32,33].
Moreover, wormholes may play an important role
 which could force the cosmological constant
to be zero [34].
The wormhole ground state may be defined by a path integral over all possible
asymptotic
Euclidian 4-geometries and matter fields whose
energy-momentum tensor vanishes at  infinity.
Excited wormhole states would have sources at infinity.
However, the question concerning the  differences between a wormhole ground
state
and the excited states may  not have  a simple answer.
In fact, if  a ground state  has been found (like in [16]) then excited states
may be obtained from the repeated aplication of operators
(like ${\partial \over {\partial \phi}}$, e.g.),
checking their regularity
 and
implementing their orthonormality. But it is another issue if we happen to
find a set of solutions from the Wheeler-DeWitt equation and try to
identify which ones correspond to a   wormhole ground state or to excited
states. Recent
investigations on this issue [28,30] claim that what may be  really
relevant is to use the whole  basis of wormhole solutions
(namely, to calculate the effects of wormhole physics from Green's functions,
where these have been factorized by introducing a {\it complete} set of
wormhole states [16])
and not just trying to identify and label a  explicit
expression which would correspond either to a wormhole  ground
state or an excited one.

The Hartle-Hawking (or no-boundary proposal) [1,10] solution is expressed in
terms
of a Euclidian path integral. It is essentially a topological statement about
the class of histories summed over. To calculate the no-boundary
wave function we are required to regard a three-surface as the {\it only}
boundary of a compact four-manifold, on which the  four-metric is
$g_{\mu\nu}$ and induces $h_{ij}^0$ on the boundary, and the matter field
is $\phi$ and matches $\phi_0$ on the boundary as well. We are then instructed
to
perform a path integral over all such $g_{\mu\nu}$ and $\phi$ within all such
manifolds.
For manifolds of the form of ${\bf R} \times \Sigma$, the no-boundary proposal
indicates us to choose initial conditions at the initial point as to ensure the
closure of the
four geometry. It basically consists in setting the initial three-surface
volume $h^{1/2}$ to
zero but also involve regular conditions on the derivatives of the remaining
components of the
three-metric and the matter fields [1,10].

Let us now  briefly exemplify
how wormhole states seem to be absent and
why a Hartle-Hawking solution is only partially       determined.
Considering the more general theory
of N=1 supergravity with supermatter [35],
 we  take
a closed FRW model with complex scalar fields
$\phi, \ol\phi$,
their  fermionic partners, $\chi_A$, $\ol\chi_{A'}$,
and   a  two-dimensional  spherically symmetric K\"ahler geometry.
 The tetrad of the
four-dimensional theory can be  simplified to be:
$$ e_{a\mu} = \pmatrix {
N (\tau) &0 \cr
0 &a (\tau) E_{\hat a i} \cr }~,
{}~ \eqno (1)$$
where $ \hat a $ and $ i $ run from 1 to 3.
$ E_{\hat a i} $ is a basis of left-invariant 1-forms on the unit $ S^3 $
with volume $ \sig^2 = 2 \pi^2 $.

This Ansatz reduces the number of degrees of freedom provided by $ e_{AA'
\mu} $. If supersymmetry invariance is to be retained, then we need an
Ansatz for $ \psi^A_{~~\mu} $ and $ \bar\psi^{A'}_{~~\mu} $ which reduces the
number of fermionic degrees of freedom, so that there is equality between the
number of bosonic and fermionic degrees of freedom. One is naturally led to
take $
\psi^A_{~~0} $ and $ \bar\psi^{A'}_{~~0} $ to be functions of time only. In
the four-dimensional Hamiltonian theory, $ \psi^A_{~~0} $ and $ \ti
\psi^{A'}_{~~0} $ are Lagrange multipliers which may be freely specified. For
this reason we do not allow $ \psi^A_{~~0} $ and
 $ \bar\psi^{A'}_{~~0} $ to depend on $ \psi^A_{~~i} $ or
$ \bar\psi^{A'}_{~~i} $ in our Ansatz. We
further take  (cf. ref. [36,37] for details)
$$
\psi^A_{~~i} = e^{AA'}_{~~~~i} \bar\psi_{A'}~, ~
\bar\psi^{A'}_{~~i} = e^{AA'}_{~~~~i} \psi_A~, \eqno (2)  $$
where we introduce the new spinors $ \psi_A $ and $ \bar\psi_{A'} $ which
are functions of time only. It is important to stress that
the reason why we throw away the spin ${3 \over 2}$ modes of
the gravitino fields is we ought to have an equal number of bosonic and
fermionic degrees of freedom, which is a necessary
condition for the supersymmetry invariance to be retained.
This is a direct consequence of assuming FRW geometry.
The scalar super-multiplet, consisting of the complex massive scalar
field $ \phi, \ol \phi $ and  spin-$\half$ field $ \rchi_A, \bar\rchi_{A'}$
are chosen to be
spatially homogeneous, depending only on time.

The main result that follows (cf. eq. (21), (24))were shown {\bf not} to depend
on the
fermionic derivative factor ordering and possible K\"ahler geometry
 [25].  For a two-dimensional spherical symmetric K\"ahler
manifold we have
$$ \rg_{\phi \bar\phi}= { 1 \over (1 + \phi \bar \phi)^{2} }
{}~,\rg^{\phi \bar\phi}=~(1 + \phi \bar \phi)^{2} ~.\eqno(3)$$
The Levi-Civita connections of the   K\"ahler manifold are
$$ \Gamma^{\phi}_{\phi \phi} = g^{\phi \bar \phi} { {\pt g_{\phi \bar \phi} }
\over {\pt\phi}} = -2 { \bar \phi \over (1 + \phi \bar \phi)}  ~\eqno(4)$$
and its complex conjugate. The rest of the components are zero.

Using the Ans\"atze described  previously,
the action of the full theory
 can be reduced to one with a
 finite number of degrees of freedom.
Starting from the action so  obtained,
 we  study the Hamiltonian formulation of this model.
The general
 Hamiltonian of a locally  supersymmetric model may be put in the simplified
form
$$ \eqalignno {
H = {\cal N} \cH + \psi^A_{~~0} S_A &+ \bar S_{A'} \bar
\psi^{A'}_{~~0} \cr
 + M_{A B} J^{A B} &+ \bar M_{A' B'} \bar J^{A' B'}~, &(5)
\cr } $$
expected for a theory with the corresponding gauge invariances. Here $\cal  N $
is the lapse function, while $ \cH $
is basically a Wheeler-DeWitt operator. $ S_A $ and $ \bar S_{A'} $ are the
local supersymmetry
generators,  and $ J^{A B} $
and $ \bar J^{A' B'} $ are the generators of local
Lorentz rotations, while $ M_{A B} $ and $ \bar M_{A' B'} $ are Lagrange
multipliers giving the amount of Lorentz rotation applied per unit time.
Classically, the
constraints vanish, and the set of (first-class)
constraints forms an algebra. In the Hamiltonian decomposition, the variables
are split into dynamical
components $ e^{A A'}_{~~~~i},~\psi^A,~\bar\psi^{A'}, ~\phi,~\bar\phi,
\rchi_{A},\bar\rchi_{A'} $,
 which together with the bosonic momenta
are the basic dynamical variables of the theory, and the
Lagrange multipliers $ {\cal N},~\psi^A_{~~0}$,$~\bar
\psi^{A'}_{~~0},$
{}~$M_{A B}$,~$\bar M_{A' B'} $ of Eq.~(5), where $\cal  N$
is formed from the $ e^{A A'}_{~~~~0} $ and the $ e^{A A'}_{~~~~i} $,
and $ M_{A B}, \bar M_{A' B'} $ involve the zero components $ \om_{A B
0}, \bar\om _{A' B' 0} $ of the connection. One computes the canonical
momenta conjugate to the dynamical variables listed above in the usual way.
The constraint generators are functions of the
basic dynamical variables. For the
gravitino and spin-$\half$ fields, the canonical momenta give second-class
constraints. These are eliminated when
Dirac
brackets are introduced [24] instead of the original Poisson brackets.

 The procedure to find the expressions of $S_{A}$ and $\bar S_{A'}$ is  simple.
 First, we have to calculate the conjugate momenta of the
 dynamical variables and then evaluate the  expression (5).
Afterwards, we  read out the coefficients of $\psi_{0}^{~A}$    and
 $\bar \psi_{0}^{~A'}$   from this expression in order to get the $S_{A}$ and
$\bar S_{A'}$
constraints, respectively.
We also  need to redefine the  $ \rchi_{A} $ field and $ \psi_{A} $ field in
order to simplify the Dirac
brackets [25]:

$$ \hat \rchi_{A} = {\sigma a^{3 \over 2} \over 2^{1 \over 4} (1 + \phi \bar
\phi)} \rchi_{A}~, ~  \hat{ \bar \rchi}_{A'} = {\sigma a^{3 \over 2} \over 2^{1
\over 4} (1 + \phi \bar \phi)} \bar \rchi_{A'} ~.\eqno(6) $$
The conjugate momenta become
$$ \pi_{\hat \rchi_{A}} = -i n_{AA'} \hat{ \bar \rchi}^{A'}~, ~
 \pi_{\hat{ \bar \rchi}_{A'}} = -i n_{AA'} \hat \rchi^{A} ~.\eqno(7) $$
This pair form a set of   second class constraints. Consequently, the Dirac
bracket ($[~]_{D}$) becomes

$$ [ \hat \rchi_{A} , \hat{\bar \rchi}_{A'} ]_{D} = -i n_{AA'}~. \eqno(8) $$
Similarly for the $ \psi_{A}$  field,

$$ \hat \psi_{A} = {\sqrt{3} \over 2^{1 \over 4}} \sigma a^{3 \over 2}
\psi_{A}~, ~
\hat{\bar \psi}_{A'} = {\sqrt{3} \over 2^{1 \over 4}} \sigma a^{3 \over 2} \bar
\psi_{A'} ~,\eqno(9) $$
where the conjugate momenta are

$$ \pi_{\hat{ \psi}_{A}} = in_{AA'} \hat{\bar \psi}^{A'} ~,~
\pi_{\hat{\bar  \psi}_{A'}} = in_{AA'} \hat \psi^{A}~.  \eqno(10)$$
The Dirac bracket is then

$$ [\hat \psi_{A} , \hat{\bar \psi}_{A'}]_{D} = in_{AA'}~. \eqno(11) $$
Furthermore,
$$ [a , \pi_{a}]_{D} = 1~, ~ [\phi, \pi_{\phi}]_{D} = 1~,
 ~[\bar \phi, \pi_{\bar \phi}]_{D} = 1~, \eqno(12)$$
and the rest of the brackets are zero.
After substituting  the redefined  fields in the  constraints,
 we drop the hat over the new variables.

It is simpler to describe the theory using only (say) unprimed spinors, and, to
this end, we define
$$ \bar \psi_{A} = 2 n_{A}^{~B'} \bar \psi_{B'}~, ~
 \bar \rchi_{A} = 2 n_{A}^{~B'} \bar \rchi_{B'} ~,\eqno(13) $$
with which the new Dirac brackets are
$$ [\rchi_{A}, \bar \rchi_{B}]_{D} = -i \epsilon_{AB}~, ~
 [\psi_{A}, \bar \psi_{B}]_{D} = i \epsilon_{AB} ~.\eqno(14) $$
The rest of the brackets remain unchanged. Using these new variables, the
supersymmetry constraints are
$$ S_{A} = {1 \over \sqrt{2}} (1 + \phi \bar \phi) \rchi_{A} \pi_{\phi} - {i
\over 2 \sqrt{6}} a \pi_{a} \psi_{A} $$
$$ - \sqrt{3 \over 2} \sigma^{2}a^{2} \psi_{A} - {5i \over 4 \sqrt{2}} \bar
\phi \rchi_{A} \bar \rchi_{B} \rchi^{B} $$
$$+{1 \over 8 \sqrt{6}} \psi_{B} \bar \psi_{A}  \psi^{B} - {i \over 4 \sqrt{2}}
\bar \phi \rchi_{A} \psi^{B} \bar \psi_{B} $$
$$ +{5 \over 4 \sqrt{6}} \rchi_{A} \psi^{B} \bar \rchi_{B} + {\sqrt{3} \over 4
\sqrt{2}} \rchi^{B} \bar \rchi_{A} \psi_{B} $$
$$ - {1 \over 2 \sqrt{6}} \psi_{A} \rchi^{B} \bar \rchi_{B}  \eqno(15a)$$
and

$$ \bar S_{A} = {1 \over \sqrt{2}} (1 + \phi \bar \phi) \bar \rchi_{A}
\pi_{\bar \phi} + {i \over 2 \sqrt{6}} a \pi_{a} \bar \psi_{A} $$
$$ - \sqrt{3 \over 2} \sigma^{2}a^{2} \bar \psi_{A} + {5i \over 4 \sqrt{2}}
\phi \bar \rchi_{B} \rchi^{B} \bar \rchi_{A} $$
$$-{1 \over 8 \sqrt{6}} \bar \psi^{B} \psi_{A} \bar \psi_{B} - {i \over 4
\sqrt{2}} \phi \psi_{B} \bar \psi^{B} \bar \rchi_{A} $$
$$ +{5 \over 4 \sqrt{6}} \rchi^{B}  \bar \psi_{B} \bar \rchi_{A} - {\sqrt{3}
\over 4 \sqrt{2}} \bar \psi_{B} \rchi_{A} \bar \rchi^{B} $$
$$ - {1 \over 2 \sqrt{6}} \rchi^{B} \bar \rchi_{B}  \bar \psi_{A} \eqno(15b)$$
Quantum mechanically, one replaces the Dirac brackets by  anti-commutators if
both arguments are odd
(O) or commutators if
otherwise (E):
$$ [E_{1} , E_{2}] = i [E_{1} , E_{2}]_{D} ~,~ [O , E] = i [O , E]_{D} ~,~
\{O_{1} , O_{2}\} = i [O_{1} , O_{2}]_{D} ~. $$
We  use  $\hbar=1$ and $\sigma^2 = 2\pi^2$.
We choose $ (\rchi_{A} , \psi_{A} , a , \phi , \bar \phi) $ to be the
coordinates and
($ \bar \rchi_{A},$ $ \bar \psi_{A}, $ $ \pi_{a},$ $ \pi_{\phi}$ ,$ \pi_{\bar
\phi} $)
to be the momentum operators. Hence
$$ \bar \rchi_{A} \rightarrow - {\pt \over \pt \rchi_{A}} ,~ \bar \psi_{A}
\rightarrow {\pt \over \pt \psi_{A}},~
 \pi_{a} \rightarrow -i{\pt \over \pt a} , ~\pi_{\phi} \rightarrow -i {\pt
\over \pt \phi},~
 \pi_{\bar \phi} \rightarrow -i {\pt \over \pt \phi}  \eqno(16)$$

Some criteria have been presented to determine a suitable factor ordering for
(15a), (15b).
This  problem  is related to the presence of cubic terms
in the supersymmetry constraints.
Basically,
$S_A, \ol S_A, {\cal H}$ could be chosen by requiring that [37,39]:

{\bf 1.}~ $S_A \Psi =0 $ describes the transformation properties of $\Psi$
under right handed supersymmetry
transformations (in the  ($a, \psi_A$) representation),

{\bf 2.}~ $\ol S_A \Psi =0 $ describes the transformation properties of $\Psi$
under left handed supersymmetry
transformations (in the  ($a, \ol \psi_A$) representation),

{\bf 3.}~ $S_A, \ol S_A$ are Hermitian adjoints with respect to an adequate
inner product [5],

{\bf 4.}~ A Hermitian Hamiltonian ${\cal H}$ is defined by consistency of the
quantum algebra.
However,  not all of these criteria can be
satisfied simultaneously (cf. [37,39]). An arbitrary choice is to satisfy {\bf
1,2,4} as in
here and [24,25,37,39,40]. Another possibility (as in [24,25,38]) is to
go beyond this factor ordering and insist that $S_A, \ol S_A$  could still
be related by a Hermitian adjoint operation (requirement {\bf 3.}). If we adopt
this
then there are some quantum corrections to $S_A, \ol S_A$ (namely, adding
terms linear in
$\psi_A, \chi_A$ to $S_A$ and linear in
$\ol \psi_A, \ol\chi_A$ to $\ol S_A$) which nevertheless modify the
transformation rules for the
wave function under supersymmetry requirements {\bf 1,2.}

Following the ordering used in ref.[24,25,37,39,40],
we  put all the fermionic derivatives in  $S_{A}$ on the right. In $\ol S_A$
all the
fermionic derivatives are on the left. Hence,

$$ S_{A} = -{i \over \sqrt{2}} (1 + \phi \bar \phi) \rchi_{A} {\pt \over  \pt
\phi}
 - {1 \over 2 \sqrt{6}} a \psi_{A} {\pt \over \pt a} $$
$$ - \sqrt{3 \over 2} \sigma^{2}a^{2} \psi_{A}
- {5i \over 4 \sqrt{2}} \bar \phi \rchi_{A} \rchi^{B} {\pt \over \pt
\rchi^{B}}$$
$$-{1 \over 8 \sqrt{6}} \psi_{B} \psi^{B} {\pt \over \pt \psi^{A}}
- {i \over 4 \sqrt{2}} \bar \phi \rchi_{A} \psi^{B} {\pt \over \pt \psi^{B}} $$
$$-{5 \over 4 \sqrt{6}} \rchi_{A} \psi^{B}{\pt \over \pt \rchi^{B}}
 + {\sqrt{3} \over 4 \sqrt{2}} \rchi^{B} \psi_{B} {\pt \over \pt \rchi^{A}} $$
$$+ {1 \over 2 \sqrt{6}} \psi_{A} \rchi^{B} {\pt \over \pt \rchi^{B}},
\eqno(17a)$$
and


$$ \bar S_{A} = {i \over \sqrt{2}} (1 + \phi \bar \phi) {\pt \over \pt
\rchi^{A}}
 {\pt \over \pt \bar \phi} + {1 \over 2 \sqrt{6}} a {\pt \over \pt a} {\pt
\over \pt \psi^{A}} $$
$$ - \sqrt{3 \over 2} \sigma^{2}a^{2} {\pt \over  \pt \psi^{A}}
+ {5i \over 4 \sqrt{2}} \phi {\pt \over \pt \rchi^{A}} {\pt \over \rchi^{B}}
\rchi^{B} $$
$$-{1 \over 8 \sqrt{6}} \epsilon^{BC} { \pt \over \pt \psi^{B}} {\pt \over \pt
\rchi^{C}} \psi_{A}
 - {i \over 4 \sqrt{2}} \phi {\pt \over \pt \psi^{B}} {\pt \over \pt \rchi^{A}}
\psi^{B} $$
$$ -{5 \over 4 \sqrt{6}} {\pt \over \pt \psi^{B}} {\pt \over \pt \rchi^{A}}
\rchi^{B}
- {\sqrt{3} \over 4 \sqrt{2}} \epsilon^{BC} {\pt\over \pt \psi^{B}} {\pt \over
\pt \rchi^{C}} \rchi_{A}$$
$$ - {1 \over 2 \sqrt{6}} {\pt \over \pt \psi^{A}} {\pt \over \pt \rchi^{B}}
\rchi^{B}
\eqno(17b)$$

The Lorentz constraint  $ J_{AB} = \psi_{(A} \bar
\psi_{B)} - \rchi_{(A} \bar \rchi_{B)} $
imply for  $\Psi$
$$ \Psi = A + iB \psi^{C} \psi_{C} + C \psi^{C} \rchi_{C} + iD \rchi^{C}
\rchi_{C} +
E \psi^{C} \psi_{C} \rchi^{D} \rchi_{D}~, \eqno (18)$$
where $A$, $B$, $C$, $D$, and $E$ are functions of $a$, $\phi$ and $\bar \phi$
only.
Using eq. (17a), (17b),   we   get four equations from  $ S_{A} \Psi = 0 $
and  another four equations from $ \bar S_{A} \Psi = 0 $ (all first order
differential equations):

$$ -{ i \over \sqrt{2}} (1 + \phi \bar \phi) {\pt A \over \pt \phi} = 0~,
\eqno(19a)$$
$$ -{ a \over 2 \sqrt{6}} {\pt A \over \pt a} - \sqrt{3 \over 2} \sigma^{2}
a^{2} A = 0~,  \eqno(19b)$$
$$ (1 + \phi \bar \phi) { \pt B \over \pt \phi} + {1 \over 2} \bar \phi B
+ { a \over 4 \sqrt{3}} {\pt C \over \pt a} - {7 \over 4 \sqrt{3}} C +
{\sqrt{3} \over 2}  \sigma^{2} a^{2} C = 0~, \eqno(19c) $$
$$ {a \over \sqrt{3}} {\pt D \over \pt a} + 2 \sqrt{3} \sigma^{2} a^{2} D -
\sqrt{3} D
- (1 + \phi \bar \phi) {\pt  C \over \pt \phi} - {3 \over 2} \bar \phi C = 0~,
\eqno(19d) $$
$$ i \sqrt{2} (1 + \phi \bar \phi) {\pt E \over \pt \bar \phi} = 0~,
\eqno(20a)$$
$$ {a \over \sqrt{6}} { \pt E \over \pt a} - \sqrt{6} \sigma^{2} a^{2} E = 0~,
\eqno(20b) $$
$$ {a \over \sqrt{3}} {\pt B \over \pt a} - 2 \sqrt{3} \sigma^{2} a^{2} B -
\sqrt{3} B
 + (1 + \phi \bar \phi) {\pt C \over \pt \bar \phi} + {3 \over 2} \phi C = 0
{}~,\eqno(20c)$$
$$  (1 + \phi \bar \phi) {\pt D \over \pt \bar \phi} + {1 \over 2} \phi D
- {a \over 4 \sqrt{3}} {\pt C \over \pt a} + {7 \over 4 \sqrt{3}} C + {\sqrt{3}
\over 2} \sigma^{2} a^{2} C = 0 ~. \eqno(20d)$$
We can see that (19a), (19b) and (20a), (20b)
constitute  decoupled equations for $A$ and $E$, respectively.
 They have the general solution
$$ A = f(\bar \phi) \exp({-3 \sigma^{2} a^{2}})~,~
 E = g(\phi) \exp ({3 \sigma^{2} a^{2}}) \eqno(21) $$
where $ f , g $ are arbitrary anti-holomorphic and
holomorphic functions of $\phi$, $\ol \phi$,  respectively.

Eq. (19c) and (19d) are coupled equations between $B$ and $C$ and eq. (20c) and
(20d) are coupled equations between $C$ and $D$.
 The first step to decouple these equations is as follows.
 Let $ B = \tilde B (1 + \phi \bar \phi)^{- {1 \over 2}} $ ,
$ C = {\tilde C  \over \sqrt{3}}(1 + \phi \bar \phi)^{- {3 \over 2}} $ , $ D =
\tilde D (1 + \phi \bar \phi)^{- {1 \over 2}}$.
Equations (19c), (19d), (20c) and (20d) then  become

$$ (1 + \phi \bar \phi)^{2} {\pt \tilde B \over \pt \phi} + {a \over 12} {\pt
\tilde C \over \pt a}
- {7 \over 12} \tilde C + {1 \over 2} \sigma^{2} a^{2}  \tilde C =
0~,\eqno(22a) $$
$$ (1 + \phi \bar \phi)^{2} {\pt \tilde D \over \pt \bar \phi} - {a \over 12}
{\pt \tilde C  \over \pt a}
+ {7 \over 12} \tilde C + {1 \over 2} \sigma^{2} a^{2}  \tilde C = 0~,
\eqno(22b) $$
$$ {\pt \tilde C \over \pt \phi} - a {\pt \tilde D \over \pt a} - 6 \sigma^{2}
a^{2} \tilde D + 3 \tilde D = 0 ~,\eqno(22c) $$
$$ {\pt \tilde C \over  \pt \bar \phi} + a {\pt \tilde B  \over \pt a} - 6
\sigma^{2} a^{2} \tilde B - 3 \tilde B = 0~. \eqno(22d)$$
 From (22a) and (22d), we can eliminate $\tilde B$ to get a partial
differential equation for $\tilde C$:
$$  (1 + \phi \bar \phi)^{2} {\pt \tilde C \over \pt \bar \phi \pt \phi}
 - {a \over 12} {\pt \over \pt a} \left(a {\pt \tilde C \over \pt a}\right)
+ {5 \over 6} a {\pt \tilde C \over \pt a} + \left[ 3 \sigma^{4} a^{4} + 3
\sigma^{2} a^{2} - {7 \over 4} \right] \tilde C = 0~, \eqno(23a) $$
and from (22b) and (22c), we will get another
partial differential equation for $\tilde C$:
$$  (1 + \phi \bar \phi)^{2} {\pt \tilde C \over \pt \bar \phi \pt \phi}
 - {a \over 12} {\pt \over \pt a} \left(a {\pt \tilde C \over \pt a}\right) +
{5 \over 6} a {\pt \tilde C \over \pt a}
 + \left[ 3 \sigma^{4} a^{4} - 3 \sigma^{2} a^{2} - {7 \over 4} \right] \tilde
C = 0~.
 \eqno(23b) $$
We can see immediately that  $\tilde C = 0$ because the coefficients of
$\sigma^{2} a^{2} \tilde C$ are different for these two equations.
 Using this result, we find (cf. ref. [25] for more details)
$$ B = h(\bar \phi) (1 + \phi \bar \phi)^{- {1 \over 2}} a^{3}
\exp({3 \sigma^{2} a^{2}})~,~
 C = 0 ~,~ D = k(\phi) (1 + \phi \bar \phi)^{- {1 \over 2}} a^{3} \exp ({-3
\sigma^{2} a^{2}} ) ~.
\eqno(24)$$
Result (24) is a direct consequence that we  could not find a consistent
(Wheeler-DeWitt type) second-order differential equation for $C$ and hence to
$B,D$.
It came directly from the corresponding first order differential equations.
Changing $S_A, \ol S_A$  in order that they can be related by some Hermitian
adjoint transformation
({\bf 3.})
gives essentially
the same outcome [25]. With a two-dimensional flat
K\"ahler geometry we get a similar result.

 While Lorentz invariance allows the pair $\psi_A\chi^A$ in (18),
supersymmetry  rejects it.
A possible interpretation could be that supersymmetry transformations
 forbid any fermionic bound state
 $\psi_A\chi^A$ by treating the  spin-$\half$ fields
 $\psi^A, \chi^B$ differently.

A Hartle-Hawking wave function\footnote{$^{1}$}{{\sevenrm
The Hartle-Hawking solution  could  not be  found
in the Bianchi-IX model of ref. [27].
Either a different homogeneity condition (as in [13]) for
$\psi^A_i$ or the framework of [8] could assist us in  this particular
problem.}}
 could be identified in the fermionic filled sector, say,
$  g(\phi) \exp ({3 \sigma^{2} a^{2}})$, but for particular expressions of
$ g(\phi)$.  We notice though that (17a), (17b), (18) are not enough to specify
$ g(\phi)$. A similar situation is also present in
ref. [38], although an extra multiplicative factor of  $a^5$ multiplying
$ g(\phi)$ induces a less clear situation. In fact, no attempt was made in ref.
[38,40]
to obtain a Hartle-Hawking wave function solution. Being $N=1$ supergravity
considered
as a square root of general relativity [3], we would expect to be able
to find solutions of the type $e^{ik\phi} e^{a^{2}}$. These would correspond
to a FRW model with a massless minimally coupled scalar field in
ordinary quantum cosmology [1,43].

In principle, there are no physical arguments for wormhole
states to be absent in N=1 supergravity with supermatter.
In ordinary FRW quantum cosmology with scalar matter fields,
the wormhole ground state solution
would have a form like $e^{-a^2\cosh(\rho)}$,
where $\rho$ stands for a matter fields function
[16,28-30]. However,
such behaviour is not
provided by eq. (24). Actually, it seems quite different. Moreover, we may
ask in which conditions can solutions (21), (24) be accomodated in order for
wormhole type solutions to be obtained.
The arbitrary functions $f(\phi, \ol\phi), g(\phi, \ol\phi),
h(\phi, \ol\phi), k(\phi, \ol\phi)$  do not allow to conclude unequivovally
that in  these  fermionic
sectors the corresponding bosonic amplitudes
would be damped at   large 3-geometries for any allowed value of
$\phi, \ol\phi$ at  infinity. Claims were then made in ref. [24,25]
that no wormhole states could be found.  The reasons were that  the Lorentz and
supersymmetry constraints do  not seem  sufficient in this case to specify
the $\phi, \ol\phi$ dependence of  $f, g, h, k$.

Hence, we  have a canonical formulation of N=1 supergravity which constitutes
a (Dirac) like square root of gravity [3,4,5].
Quantum wormhole and
Hartle-Hawking solutions were found in minisuperspaces for
pure N=1 supergravity [8,11-15,18,19,36,37,39] but
the former state is absent in the literature
\footnote{$^{2}$}{{\sevenrm
Notice that for pure gravity neither classical or quantum wormhole
solutions have been produced in the literature. A matter field seems to be
required: the ``throat'' size is proportional
to $\sqrt{\kappa}$ where $\kappa$ represents   the (conserved) flux of
matter fields.}},
for pure gravity cases
[1,9,16,28-30]. Hartle-Hawking  wave functions and wormhole ground states
are present in
ordinary minisuperspace with matter
[1,9,16,28-30]. When supersymmetry is introduced [24-27,37-39]
we face some problems
within the more general theory of N=1 supergravity with
supermatter [35]  (cf. ref. [25-27]) as far as Hartle-Hawking or wormhole type
solutions are concerned.
An attempt [40] using the constraints present in [37,39] but the
ordering employed above, also  seemed to have failed in getting wormhole
states. In addition, a model
combining a conformal scalar field with spin-$\half$ fields
(expanded in  spin$-\half$ hyperspherical harmonics and
integrating over the spatial coordinates [32]) did not produce any
wormhole solution as well [41].
However, ref. [38] clearly represents an opposite point of view, as it
explicitly depicts wormhole ground states in a locally supersymmetric setting.

It might be interesting  to point  that the
constraints employed in [38]
(and also in [37,39,40]) were derived from a {\it particular}
model constructed in [42], while ours [25] come directly from
the {\it more general} theory of N=1 supergravity coupled to
supermatter [35].
 Moreover, there are many differences
between the expressions in [36-39] and the one  hereby (see also [25]),
namely on numerical coefficients.

Let me sketch briefly how the supersymmetry constraints expressions in
[38] were obtained. First, at the pure N=1 supergravity level,
the following re-definition of fermionic non-dynamical variables
$$ \rho^A \sim a^{-1/2}\psi^A_0 +
{\cal N} a^{-2} n^{AA'}\overline{\psi}_{A'}, \eqno(25)$$ and its hermitian
conjugate
were introduced for a FRW model, changing the supersymmetry and Hamiltonian
constraints.
As a consequence, no
fermionic terms were present in  ${\cal H} \sim \{S_A, \ol S_A\}$ and no
cubic fermionic terms in the supersymmetry constraints. Hence,
no ordering problems with regard to fermionic derivatives were present. The
model with matter
was then extracted {\it post-hoc} [37,39] from a few basic assumptions
about their general form and supersymmtric algebra. This simplified route
seemed
to give similar expressions, up to minor field redefinitions,
to what we would obtain for a reduced model from the {\it particular}
theory presented in [42], as stated in
[37,39]. Note that cubic fermionic terms
like $\psi\ol\psi\psi$ or $\psi\ol\chi\chi$ are now present but the former is
absent
in the pure case.
In ref. [37,39,40], criteria {\bf 1,2,4} were used for the fermionic derivative
ordering, while
in ref. [38] one has  insisted to accomodate an Hermitian adjoint relation
between the supersymmetry transformations ({\bf 3.}). It so happens that a
wormhole
ground state was found in the former but not in the latter. In ref. [25]
the same possibilites for using these criteria  were employed but
with supersymmetry and Hamiltonian cosntraints  directly obtained from
$\psi^A_0, \ol \psi^{A'}_0, {\cal N}$ (see eq. (25)). Apparently, no wormhole
states were present. Moreover, we also recover a solution which satisfied only
partially
the no-boundary proposal conditions (see eq. (21)). A similar but yet less
clear
situation also seems to be present in ref. [38].

The issue concerning the existence or not of  wormhole and
Hartle-Hawking quantum cosmological states for
 minisuperspaces within N=1 supergravity with supermatter is therefore of
relevance [26].
The current literature on the subject is far from a consensus.
No explanation has been provided for the
(apparent) opposite conclusions in [25,38] concerning the
existence of wormhole   states and to point out which is right and why.
Furthermore, it does not seem possible for
the procedure presented  in [8] to solve
this conundrum.

Here  an answer for this particular problem is presented.
The explanation   is that chosing the choice  of Lagrange multipliers
and  fermionic derivative ordering we use can
 make a
difference. Our arguments are as follows.

On the one hand, the quantum formulation of wormholes in ordinary
quantum cosmology has been shown to depend on the lapse function [29,30].
Such ambiguity has already been pointed out
in [43] (see also [44]) but for generic quantum cosmology and related to
bosonic factor ordering questions in the Wheeler-DeWitt operator.
An ordering is necessary in order to make predictions. A proposal was made
that the kinetic terms in the Wheeler-DeWitt operator should be the Laplacian
in the natural (mini)superspace element of line, i.e., such that it would
be invariant under changes of coordinates in minisuperspace [43]. Basically,
this includes the  Wheeler-DeWitt operator to be locally self-adjoint in the
natural
measure generated by  the above mentioned element of line.
However, it suffers from the problem that the connection defined by a
minisuperspace
line element like $ds^2 = {1 \over {\cal N}} f_{\mu\nu} dq^\mu dq^\nu$ could
not be linear
on $\cal N$. This would then  lead to a Wheeler-DeWitt operator {\it not}
linear in $\cal N$
as it would be  in order that $\cal N$ be interpreted as a Lagrange multiplier
(it was also proposed in ref. [43] that this possible non-linearity dependence
on $\cal N$ could cancel out in theories like supergravity where bosons and
fermions
would be in equal number of degrees of freedom).
For each choice of $\cal N$, there is a different metric in minisuperspace, all
these
metrics being related by a conformal transformation [45]. Therefore, for each
of these
choices, the quantization process will be different. In fact, for a
minisuperspace consisting
of a FRW geometry and homogeneous scalar field, a conformal coupling allows a
more
general class of solutions of the Wheeler-DeWitt equation than does the
minimally
coupled case, even if a one-to-one correspondence exists between bounde states
[45].

For some choices of $\cal N$ the quantization are
 even inadmissible,
e.g,
when ${\cal N} \rta 0$ too fast for vanishing
3-geometries in the wormhole case (cf. ref. [29,30] for more details).
Basically, requiring regularity for $\Psi$ at $a\rta 0$ is equivalent
to self-adjointness for the Wheeler-DeWitt operator at that point. Such
extension
would be expected since wormhole wave functions calculated via a path integral
are
regular there. Three-geometries with zero-volume would be a consequence of the
slicing procedure which has been carried. In other words, $a=0$ simply
represents a coordinate singularity in minisuperspace. An extension
for (and beyond it), similar to the case of the Rindler wedge and the full
Minkowski space,
would be desirable. The requirement that the Wheeler-DeWitt operator be
self-adjoint
selects a scalar product and a measure in minisuperspace. Gauge choices of
$\cal N$
that vanish too fast when $a \rta 0$ will lead to problems as the
minisuperspace measure will be infinite at (regular) configurations associated
with vanishing three-geometries volume.
The difference on the quantization
manifests itself in the Hilbert space structure of the wormhole
solutions [29-31] due to the scalar product dependence on $\cal N$ and not in
the structure of
the Wheeler-DeWitt operator or path integral.
More precisely, the formulation of
global laws, i.e., finding boundary conditions
for the Wheeler-DeWitt equation in the wormhole case, equivalent
to the ones in the path integral approach,
could depend
on the choice of $\cal N$ but not the
local laws in minisuperspace\footnote{$^{3}$}{{\sevenrm
Physical results such as effective interactions are independent of the
choice of $\cal N$ due to the way the corresponding path integrals  are
formulated.}}.

On the other hand, a similar effect seems to occur when local supersymmetry
transformations
are present. Besides the lapse function, we have now the
time components of the gravitino field, $\psi^A_0$, and
of the torsion-free connection $\omega_{AB}^0$ as Lagrange multipliers.
If we use transformation (25) but without the last term, then
the supersymmetry and Hamiltonian constraints read  (in the pure case):

$$ S_A = \psi_A\pi_a - 6ia \psi_A +
{{i}\over{2a}}n_A^{E'}\psi^E\psi_E\overline{\psi}_{E'}, \eqno(26a)$$

$$ \overline{S}_{A'} = \overline{\psi}_{A'} \pi_a +
6ia \overline{\psi}_{A'} - {{i}\over{2a}}n_E^{A'}\overline{\psi}^{E'}
\psi_E\overline{\psi}_{E'},  \eqno(26b)$$

$$ {\cal H} = -a^{-1} (\pi_a^2 + 36 a^2)       + 12a^{-1} n^{AA'}\psi_A
\overline{\psi}_{A'}. \eqno(26c)$$

If $\rho_A, \ol \rho_{A'}$ had been used instead of $\psi^A_0, \ol \psi^{A'}_0$
then the
second terms in (26a)-(26c) would be absent. I.e.,
for the transformation  (25) the corresponding supersymmetry constraints and
the Hamiltonian are
either linear or free of fermionic terms. What seems to
have been gone  unnoticed is the following.
{\it Exact}  solutions of $S_A \Psi =0$ and $\ol S_A \Psi =0$ (using the
criteria {\bf 1,2,4})
in the pure case for (26a),(26b) with or without second term
are  $A_1 = e^{-3a^2}$ and $A_2 = e^{3a^2}$, respectively, for
$\Psi = c A_1 + d A_2 \psi_A\psi^A$ where $c, d$ are constants.
This $\Psi$ also constitutes a linear combination of WKB solutions of
${\cal H}\Psi =0$, obtained form the corresponding Hamilton-Jacobi equation,
i.e.,
they represent a {\it semi-classical approximation},
{\it but only} for the $\cal H$ without the second term in (26c), i.e.,
when (25) is fully employed. In fact,  it does not for the full expression in
(26c);  the function $e^{3a^2}$ would have to be replaced.

Hence the choice between $\rho_A$ and $\psi^A_0$ directly affects any
consistency between the
quantum solutions of the constraints (26a)-(26c). Moreover, an important point
(which will be stressed later) is that the Dirac-like equations in ref. [38]
lead consistently to a set of Wheeler-DeWitt equations (like in [37,39,40])
but that could not be entirely achieved in ref. [25].
As explained in eq. (24), the difficulty in determining the
$\phi, \ol\phi$ dependence of $f,g,h,k$ (and therefore to acess on the
existence of wormhole states) is related to the fact that $C=0$, which
is an indication as well that  corresponding Wheeler-DeWitt equations or
solutions
could not be obtained from the supersymmetry constraints.

Choosing (25) we achieve the simplest form for
the supersymmetry and Hamiltonian constraints and their
Dirac brackets. This is important at the pure case level,
as far as the solutions of  $S_A \Psi =0$ and $\ol S_A \Psi =0$
are concerned. Moreover, fermionic factor ordering become absent in that case.
If we try to preserve this property through a {\it post-hoc}
approach [37,39] when going to the matter case (keeping a simplified
form for the constraints and algebra) then we might hope to avoid any problems
like
the ones refered  to in eq. (24). In addition, using the fermionic ordering of
[38]
where we accomodate the Hermitain adjointness with requirements {\bf 1,2,4} up
to minor
changes relatively to {\bf 1,2}, we do get a wormhole groud state. Thus,
there seems to be a relation between a choice of Lagrange multipliers (which
simplifies the constraints and the algebra in the pure case), fermionic factor
ordering
(which may become absent in the pure case) and obtaining  second order
consistency equations (i.e.,
 Wheeler-DeWitt type equations) or solutions from the
supersymmetry constraints. The failure of this last one  is the
reason why $C=0$ and $f,g,h,k$ cannot be determined from the algebra.
Different choices of $\psi^A_0$ or $\rho_A$, then of fermionic derivative
ordering will lead to
different supersymmetry constraints and to different solutions for the
quantization of the problem.
It should also be stressed that from the supersymmetric algebra a combination
of
two supersymmetry
transformations, generated by $S_A$ and $\ol S_{A'}$ and whose amount is
represented
by the Lagrange multipliers $\psi^A_0, \ol \psi^{A'}_0$, will be (essentially)
equivalent to a
transformation generated by
the Hamiltonian constraint and where the lapse function is the corresponding
Lagrange multiplier.

So, how should the search for wormholes ground
states\footnote{$^{4}$}{{\sevenrm
Regarding the Hartle-Hawking solutions  it seems it can be obtained
straightforwardly either
up to a specific definition of homogeneity [13] or following the approach in
[8]. This might help
in regarding the results found in [27] with respect to the Hartle-Hawking
solution.}}
in N=1 supergravity be addressed?
One possibility would be to employ  a
transformation like (25) (see [37]). In fact, using it from the begining in our
case model it will
change some coefficients in the supersymmetry constraints as it can be
confirmed.
As a consequence, we are then allowed to get consistent second order
differential equations
from $S_A \Psi =0$  and $\ol S_{A'} \Psi =0$. Hence, a line equivalent
to the one followed in ref. [38] can be used and a wormhole ground state be
found.
Alternatively, we could
restrict ourselves
to the {\it post-hoc} approach introduced and
followed throughout in [37,39] as explained above. Another possibility, is to
extend the approach introduced by L. Garay [28-31] in ordinary quantum
cosmology
to the cases where local supersymmetry is present.
The basic idea is that what is really relevant is to determine a {\it whole}
basis
of wormhole solutions of  the associated Wheeler-DeWitt equations,
not just trying to identify  one single solution like the ground state
from the all set of solutions. Hence, we
ought to adequatly define what a basis of wormhole solutions means.
In this case, we could be able to still use any Lagrange multiplier (just as
$\psi^A_0$),
avoiding having to find a  redefinition of fermionic variables  as in (25) but
for the matter case in question (scalar, vector field, etc).

Basically, improved boundary conditions for wormholes can be formulated
by requiring square integrability in the {\it maximmaly extended}
minisuperspace
[28,29]. This condition ensures that $\Psi$ vanishes at the {\it truly}
singular configurations
and guarantees its regularity at any other (coordinate) one,
including vanishing 3-geometries.
A   maximally extended minisuperspace and a proper
definition of its boundaries in order to
comply with  the behaviour of $\Psi$ for $a\rta 0$ and
$a\rta \infty$ seems to be mandatory  in ordinary
quantum cosmology. The reason was that the
quantum formulation of wormholes has been shown to depend on
the lapse function, $\cal N$ [28,30].
The maximal analytical extension of minisuperspaces can be considered as the
natural configuration space for quantization. The boundary of the
minisuperspace
would then consist of all those configurations which are truly singular.
Any regular configurations will be in  its interior.
Another reason to consider the above boundary conditions in a
maximally extended minisuperspace is that it allow us to avoid boundary
conditions at
$a=0$ to guaratee the self-adjointness of the Wheeler-DeWitt operator.
This operator is hyperbolic and well posed boundary conditions
can only be imposed on its characteristic surfaces and the one associated with
$a=0$ may not be of this type, like in the case of a conformally coupled scalar
field.
In such a case, it would be meaningless to require self-adjointness there
(cf. ref. [28,30] for more details).

Within  this framework wormhole solutions would form a Hilbert space.
These ideas  must then be extended    to
a case of locally supersymmetric minisuperspace with
odd Grassmann (fermionic) field variables.
In this case, not only we have to deal with different
possible behaviours for  $\cal N$ but also with $\psi^A_0$.
Then, it will be possible to determine
explicitly the form of $f,g,h,k$ in order that some or even an overlap of them
could provide
 a wormhole wave function behaviour,
including the ground state. In fact, this would mean that not only
the bosonic amplitudes $A,B,..$ would
have to be considered for solutions
but the fermionic pairs ought to be taken
as well.
Constructing an adequate Hilbert space from (24) would
lead us to a basis of wormhole states in such a
singularity-free space(see ref. [28]). Wormhole wave functions
could be interpretated in terms of overlaps between
different states.

Another point which might be of some relevance
is the following [30]. The  evaluation of the path integral
(or say, determining the boundary conditions for the Wheeler-DeWitt equation)
for
wormhole states in ordinary minisuperspace quantum cosmology requires the
writing
of an action adequate to asymptotic Euclidian space-time, through the inclusion
of
necessary boundary terms [16,28-30]. There may
changes when fermions and supersymmetry come
into play. A {\it different} action\footnote{$^{5}$}{{\sevenrm
The canonical form of action of
pure N=1 supergravity present in the literature [5]
(which includes boundary terms)
 is  not
invariant under supersymmetry transformations.
Only recently a fully invariant action but restricted to
 Bianchi class A models was presented [15].}} would then
induces improved  boundary conditions for the intervening
fields
as far a wormhole Hilbert space structure is concerned
in a locally supersymmetric minisuperspace.

Summarizing,  the issue
 concerning the existence
of wormhole states in locally supersymmetric minisuperspace
models was addressed in this work.
Wormhole  states
are apparently  absent in models obtained from
the more  general theory of N=1 supergravity with supermatter.
As explained, the cause
 investigated here is that an appropriate  choice of
Lagrange multipliers and fermionic derivative makes a difference.
 From the former we get the simplest form of the supersymmetry and
Hamiltonian constraints and their Dirac brackets in the pure case. This ensures
no
fermionic derivative ordering problems and that the solutions of the quantum
constraints are consistent.
Either from a {\it post-hoc} approach (trying to extend the  obtained framework
in the pure case)
or from a direct dimensional-reduction we  get consistent second order
Wheeler-DeWitt type
equations or corresponding solutions     in the supermatter case.
 From an adequate use of criteria {\bf 1,2,3,4} above, we get a wormhole ground
state.
We also notice  that the  use of appropriate Lagrange multipliera also requires
a specific
fermionic ordering results in order  to
 obtain a consistency set of  Wheeler-DeWitt equations or respective solutions.
The search for wormhole solutions could also be addressed from another
point of view [28,30]. One has to invest supersymmetric quantum wormholes
with a Hilbert space structure, associated with
a maximal analytical extension of the corresponding
minisuperspace.
A basis of wormhole states might  then be obtained from the many
possible solutions of the supersymmetry constraints  equations.

\medskip \medskip
\noindent
{\bf ACKNOWLEDGEMENTS}

The author is  thankful to
A.D.Y. Cheng, L. J. Garay,  S.W. Hawking and O. Obregon for  helpful
conversations
and for sharing their points of view.
The author also   gratefully acknowledges
questions and discussions with I. Antoniadis,
G. Esposito, B. Ovrut, D. Page, D. Salopek,  K. Stelle, A. Zhuk at the
VI Moskow Quantum Gravity International Semimar,
which  motivated further improvements in the paper.
This work was
supported  by
a Human Capital and Mobility (HCM)
Fellowship from the European Union (Contract ERBCHBICT930781).

\noindent
{\bf REFERENCES}

{\rm

\advance\leftskip by 4em
\parindent = -4em

[1] See for example, G. Gibbons and S.W. Hawking, {\it Euclidian Quantum
Gravity},
World Scientific (Singapore, 1993);

J. Halliwell, in: {\it Proceedings of the Jerusalem Winter School on
Quantum Cosmology and Baby Universes}, edited by T. Piram et al, World
Scientific, (Singapore, 1990)
 and refereces therein.

[2]  P. van Nieuwenhuizen, Phys.~Rep. {\bf 68} (1981) 189.

[3] C. Teitelboim, Phys.~Rev.~Lett.~{\bf 38}, 1106 (1977), Phys. Lett. B{\bf
69} 240 (1977).

[4] M. Pilati, Nuc. Phys. B {\bf 132}, 138 (1978).

[5] P.D. D'Eath, Phys.~Rev.~D {\bf 29}, 2199 (1984).

[6] G. Esposito, {\it Quantum Gravity, Quantum Cosmology and Lorentzian
Geometries},
Sprin\-ger Verlag (Berlin, 1993).

[7] S.W. Hawking, Phys. Rev. D{\bf 37} 904 (1988).

[8]  R. Graham and A. Csord\'as, Phys. Rev. Lett. {\bf 74} (1995) 4129;

R. Graham and A. Csord\'as, {\it Nontrivial Fermion States in Supersymmetric
Minisuperspace}, talk
presented at Mexican School in Gravitation and
           Mathematical Physics, Guanajuato, Mexico, Dec 12-16, 1994,
gr-qc/9503054

[9] R. Graham and A. Csord\'as, {\it Quantum states on supersymmetric
minisuperspace with a cosmological
 constant}, gr-qc/9506002

[10] J.B. Hartle and S.W. Hawking, Phys.~Rev.~D {\bf 28}, 2960 (1983).

[11] P.D. D'Eath, S.W. Hawking and O. Obreg\'on, Phys.~Lett.~{\bf 300}B, 44
(1993).

[12] P.D. D'Eath, Phys.~Rev.~D {\bf 48}, 713 (1993).

[13] R. Graham and H. Luckock, Phys. Rev. D
{\bf 49}, R4981 (1994).

[14] M. Asano, M. Tanimoto and N. Yoshino, Phys.~Lett.~{\bf 314}B, 303 (1993).

[15] H. Luckock and C. Oliwa,
(gr-qc 9412028), accepted  in
Phys. Rev. D.

[16] S.W. Hawking and D.N. Page, Phys.~Rev.~D {\bf 42}, 2655 (1990).

[17] P.D. D'Eath, Phys. Lett. B{\bf 320}, 20 (1994).

[18]  A.D.Y. Cheng, P.D. D'Eath and
P.R.L.V. Moniz, Phys. Rev. D{\bf 49} (1994) 5246.

[19]  A.D.Y. Cheng, P. D'Eath and
P.R.L.V. Moniz,
{\rm Gravitation and Cosmology}
{\bf 1} (1995) 12

[20] S. Carroll, D. Freedman, M. Ortiz and
D. Page, Nuc. Phys. B{\bf 423}, 3405 (1994).

[21] J. Bene and R. Graham, Phys. Rev. {\bf D49} (1994) 799; R. Mallett, Class.
Quantum Grav. {\bf 12} (1994) L1; A. Cheng, O. Obregon and P. Moniz, in
preparation.

[22]  A.D.Y. Cheng, P.D. D'Eath and
P.R.L.V. Moniz, DAMTP R94/44, Class. Quantum Grav. {\bf 12} (1995) 1343.

[23] P. Moniz, {\it Physical States in a Locally Supersymmetric FRW model
coupled
to Yang-Mills fields}, DAMTP Report, in preparation

[24]  A.D.Y. Cheng, P.  D'Eath and
P.R.L.V. Moniz,
{\rm Gravitation and Cosmology}
{\bf 1} (1995) 12

[25]  A.D.Y. Cheng and P.R.L.V. Moniz,
Int. J. Mod. Phys. {\bf D4}, No.2 April (1995) - to appear.

[26] P. Moniz, {\it The Case of the Missing Wormhole State},
talk presented in the 6th Moskow Quantum Gravity Seminar, 12-19 June 1995,
Moskow, Russia (to appear in the Proceedings - World Scientific), gr-qc/9506042

[27]   P. Moniz,
{\it Back to Basics?... or How can supersymmetry be used in simple quantum
cosmological model},
DAMTP R95/20, gr-qc/9505002, Communication presented at the First Mexican
School in
 Gravitation and Mathematical Physics, Guanajuato, Mexico, December 12-16,
 1994;

{\it Quantization of the Bianchi type-IX
model in N=1  Supergravity in the
presence of supermatter}, DAMTP report R95/21, gr-qc/9505048, submited to
International
Journal of Modern Physics {\bf A}

[28] L. Garay, Phys. Rev. D{\bf 48 } (1992) 1710.

[29] L. Garay, Phys. Rev. D{\bf 44} (1991) 1059.

[30] L. Garay, Ph. D. Thesis (in spanish) Madrid - Consejo Superior de
Investigaciones
Cientificas, 1992.

[31] G. Mena-Marugan, Class. Quant. Grav. {\bf 11} (1994) 2205; Phys. Rev.
D{\bf 50} (1994) 3923.

[32] A.Lyons, Nuc. Phys. {\bf B}324 (1989) 253.

[33]  H.F. Dowker , Nuc. Phys. {\bf B}331 (1990) 194;
 H.F. Dowker and R. Laflamme, Nuc. Phys. {\bf B}366 (1991) 201.

[34] S. Coleman, Nuc. Phys. {\bf B}310 (1988) 643.

[35]  J. Wess and J. Bagger, {\it Supersymmetry and Supergravity},
2nd.~ed. (Princeton University Press, 1992).

[36] P.D. D'Eath and D.I. Hughes, Phys.~Lett.~{\bf 214}B, 498 (1988).

[37] P.D. D'Eath and D.I. Hughes, Nucl.~Phys.~B {\bf 378}, 381 (1992).

[38] L.J. Alty, P.D. D'Eath and H.F. Dowker, Phys.~Rev.~D {\bf 46}, 4402
(1992).

[39] D.I. Hughes, Ph.D.~thesis, University of Cambridge (1990), unpublished.

[40] P. D Eath, H.F. Dowker and D.I. Hughes, {\it Supersymmetric
Quantum Wormholes States} in: Proceedings of the Fifth Moskow Quantum Gravity
Meeting, ed. M. Markov, V. Berezin and V. Frolov, World Scientific (Singapore,
1990).

[41] H.F. Dowker, Ph.D. Thesis, chapter 4, University of Cambridge (1991),
unpublished

[42] A. Das. M. Fishler and M. Rocek,  Phys.~Lett.~B {\bf 69}, 186 (1977).

[43] S.W. Hawking and D. Page, Nuc. Phys. B{\bf 264} (1986) 185.

[44] T. Padmanabhan, Phys. Rev. Lett. {\bf 64} (1990) 2471.

[45] D. Page, J. Math. Phys. {\bf 32} (1991) 3427.

\ \ \ }

\bye
\end